Title: IONIC CONDUCTIVITY OF SOLID OXIDES HxAg1-xTaWO6.nH2O: MICROSTRUCTURAL ASPECTS

Article Type: Research Paper

Corresponding Author: Dr. Daniel Valim, Ph.D.

Corresponding Author's Institution:

First Author: Daniel Valim, Ph.D.

Order of Authors: Daniel Valim, Ph.D.; Antônio G Souza, Ph.D.; Josue M Filho, Ph.D.; Oswaldo L Alves, Ph.D.; Eder N Silva, Ph.D.; Marco Aurelio C de Santis, Ms; Romildo J Ramos, Ph.D.

Abstract: A study of impedance spectroscopy was done for the electrical characterization of pyrochlore materials. The experiment consisted of measurements of the dependence of the impedance of such systems with the temperature during heating (25 to 110 °C) and cooling (110 to 25 °C). The goal was to try to avoid the effect of humidity in the impedance spectrum. However, for the sample with $x = 0.20$, we get an anomalous response of the conductivity during the cooling process. To explain this result, we performed micro-structure characterization in this sample and in the samples with $x = 0.80$ and $0.67$. The result is that the sample with $x = 0.20$ is well characterized by the model "easy-path'' because it has high porosity while the other two must have phase precipitate structure type Suzuki. These results served to explain why the anomaly on the Arrhenius plot occurred only in the 0.20 concentration sample: microstructure difference. In this study we can also note that the doped samples had a dependence of activation energies related to the rates of concentration. Except for the sample with concentration rate of 0.20, the activation energies during cooling are among the values of 0.60 and 0.23 eV of their pure counterparts which are registered in the literature.

Research Data Related to this Submission
--------------------------------------------------
There are no linked research data sets for this submission. The following reason is given:
The authors do not have permission to share data

# IONIC CONDUCTIVITY OF SOLID OXIDES $H_xAg_{1-x}TaWO_6 \cdot nH_2O$: MICROSTRUCTURAL ASPECTS

**D Valim** - Depto. de Matemática - UNEMAT/ Sinop, Brazil
 e-mail: daniel.valim@unemat.br
- **AG Souza Filho** - Depto. de Física - UFC/ Fortaleza, Brazil
- **JM Filho** - Depto. de Física - UFC/ Fortaleza, Brazil
- **OL Alves** - Depto. de Química - UNICAMP/Campinas, Brazil
- **MAC de Santis** - Depto. de Química - UNICAMP/Campinas, Brazil
- **EN Silva** - Depto. de Física – Universidade Federal do Maranhão/ São Luíz, Brazil
- **RJ Ramos** - Instituto de Física – Universidade Federal do Mato Grosso/ Cuiabá, Brazil

*Abstract*: *A study of impedance spectroscopy was done for the electrical characterization of pyrochlore materials. The experiment consisted of measurements of the dependence of the impedance of such systems with the temperature during heating (25 to 110 °C) and cooling (110 to 25 °C). The goal was to try to avoid the effect of humidity in the impedance spectrum. However, for the sample with x = 0.20, we get an anomalous response of the conductivity during the cooling process.*

*To explain this result, we performed micro-structure characterization in this sample and in the samples with x = 0.80 and 0.67. The result is that the sample with x = 0.20 is well characterized by the model "easy-path" because it has high porosity while the other two must have phase precipitate structure type Suzuki. These results served to explain why the anomaly on the Arrhenius plot occurred only in the 0.20 concentration sample: microstructure difference. In this study we can also note that the doped samples had a dependence of activation energies related to the rates of concentration. Except for the sample with concentration rate of 0.20, the activation energies during cooling are among the values of 0.60 and 0.23 eV of their pure counterparts which are registered in the literature.*

## INTRODUCTION

The group of pyrochlore is chemically diverse. Hogarth cited by Chelgani[1] defined the formula of pyrochlore oxides as $A_{2-m}B_2O_6 \phi_{1-n} \cdot pH_2O$ where A represents an element mono-, di-, or trivalent, B represents an element tri-, tetra-, penta-, or hexavalent and $\phi$ =O, OH, F. The structures of pyrochlore may vary, depending of the cation distribution in sites A and anions in sites $\phi$, as normal, ideal, defective and reverse. For normal pyrochlore there is a regular distribution of atoms on the A sites (at positions 16d, Wyckoff notation) and $\phi$ (at positions 18b). A typical pyrochlore has an intermediate structure between the ideal and defective structures, which are set to m = 0 (complete A cation) and m = 2 (absent A cation). Some of the normal pyrochlore's related structures show an inverse distribution of A cations and site ϕ vacancies, namely the A cations (with m <1) occupy positions 8b while the vacancies occupy the positions 16d.

The compounds under investigation in this work are of type m = 1, n = 1, p = 0, 0.5 or 1. Many pyrochlores with the general chemical formula AB'B"O$_6$, as the HTaWO$_6$ [2] has random distribution of cations B' and B" in the center of the octahedrons. These compounds crystallize in the space group Fd3m(O$_h^7$) and belong to a class of materials with 450 synthetic compositions with properties that include: catalysis, piezoelectricity, iron and iron magnetism, luminescence and colossal magnetoresistance [3,4]. Other pyrochlores exhibit another collection of interesting physical properties such as semiconducting and superconductivity [5,6]. In this structure the atoms B and X form a matrix of octahedral (BX$_6$) connected by vertices while Y and A form another array of tetrahedral (YA$_4$) sharing vertices. This structure is analogous to other types of structures such as those of the ReO$_3$ and perovskite CaTiO$_3$. However, in the pyrochlore the unit skeleton (BO$_6$)n is more complex and the cavities left by them are more open than in the other two cases. These cavities have an average diameter between 3.8 and 4.6 Å[7].

The defective pyrochlores belong to a class of cationic materials that have been extensively studied in the last years, such as in the search for good solid ionic conductors for technological applications [8] as for model systems to study ion conduction[9]. For this reason, the characterization technique used in this paper has been Impedance Spectroscopy. Impedance spectroscopy (IS) is a versatile, non-destructive and non-evasive technique, able to both electrical measurements and microstructural characterizations. For the ionic conductions mechanism studies, this technique is very useful to understand it. Because the microstructure of the material is easily described by electrical models used to fit the electrical curves obtained by IS technique.

The correct interpretation about IS variables depend on the correct previous model. In a crystalline idealized solid, the model for both impedance and permittivity is the Debye one [10]. However, in polycrystalline solids, where there is presence of micro or nano crystals, some effects associated with the material roughness, the grain borders or simply distribution in activation energies in ionic monocrystalline conductors appear [11]. In the case of polycrystalline materials, such as those studied in this work, the CPE effect is associated to the microstructure (microcrystals and borders of microcrystals).

In order to characterize materials by impedance spectroscopy, we use equivalent circuit models, where the elements represent physical processes that occur in a material, and which together allows us to understand and classify the physical properties of the studied material. In this work, we have shown the results of ionic conduction studies through impedance spectroscopy characterization of H$_x$Ag$_{1-x}$TaWO$_6$.nH$_2$O pyrochlore samples, take into account both intrinsic and extrinsic mechanisms.

# 1 EXPERIMENTAL PROCEDURE
## 1.1 Sample Synthesis

The pyrochlore precursor $KTaWO_6$ samples were synthesized via solid-state reactions from stoichiometric amounts of the reactants. The reactants were $K_2CO_3$(Riedel – 99%), $Ta_2O_5$ (Alfa – 99%) and Tungsten oxide $WO_3$ (Aldrich) with 99 % of purity. The reagents and their both manufacturer and purity is shown in **Table 1**.

The reactions can be viewed as follows:

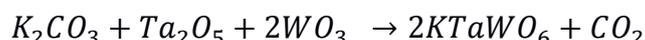

$$K_2CO_3 + Ta_2O_5 + 2WO_3 \rightarrow 2KTaWO_6 + CO_2$$

The procedure is an adaptation of procedure described by Mari[12]. The reagents were crushed and mixing in an agate mortar for 10 minutes. The system remained under heating for 45 hours, and every 15 hours the reaction was stopped for crushing and homogenization of the mixture.

**Table 1.** Characteristics of used reagents.

| Reagent | Formula | Manufacturer | Purity(%) |
|---|---|---|---|
| Tantalum Dioxide | $Ta_2O_5$ | Alfa | 99 |
| Tungsten Oxide | WO3 | Aldrich | 99 |
| Potassium Carbonate | $K_2CO_3$ | Riedel | 99 |
| Silver Nitrate | $AgNO_3$ | Merck | 99.8 |
| Nitric Acid | $HNO_3$ | Aldrich | 65* |
| Hydrochloric Acid | HCl | Merk | 36.5-38* |

The pyrochlore $HTaWO_6.H_2O$ compound was obtained by ionic exchange reaction between $KTaWO_6$ and a 12 mol.L$^{-1}$ HCl solution. The system was left under stirring at room temperature for 48 hours, with two exchanges of the supernatant solution after centrifugation of the solid.

The $HTaWO_6$ was submitted at new ionic exchange reaction. The proton ion was exchanged by $Ag^+$ to make $AgTaWO_6$. The reactions were performed using the procedures of LQES laboratory for similar compounds [13,14]. In this part of work, the reactions were optimized for a maximum exchange proton-silver. The compounds were dipping in $AgNO_3$ solution with a concentration of $10^{-3}$ at $10^{-2}$ mol.L$^{-1}$ and stirring by magnets at room temperature during 48 hours.

Finally, the protonated compounds were again subjected to ion exchange reactions, this time the proton was replaced by $Ag^+$ ions. The reactions were performed using the procedures of LQES laboratory for similar compounds [13,14]. For this time, the reactions were performed by molar rations of 4:1, 2:1, 1:1, 1:2, 1:4 and 1:8 to form $H_{1-x}Ag_xTaWO_6.H_2O$.

The compounds were immersed in aqueous $AgNO_3$ solutions and magnetically stirred at room temperature for 48 hours. The solids were then separated by centrifugation and washed with deionized water. The washing and purification process was repeated three times.

Our work consisted of a similar study of reference [15], where the $HTaWO_6$ pyrochlore samples were submitted substitutional homovalent doping with silver by an ion exchange reaction, i.e., the proton of the compound was replaced by $Ag^+$ to form compounds of type $H_{1-x}Ag_xTaWO_6$ [16]. The reactions were performed using the procedures of LQES laboratory for similar compounds [13,14]. The molar ratio H:Ag of each of the samples were 4:1, 2:1, 1:1, 1:2 and 1:4, the code samples are

HAg1:0.25, HAg1:0.5 , HAg1:1, HAg1:2 and HAg1:4. We know the ion exchange reaction, to 4:1(Hag1:0.25) for example, occurs as follows

$$4HTaWO_6 + 1AgTaWO_6 \rightarrow 5H_{0.8}Ag_{0.2}TaWO_6$$

Where we can adjust the x, which is the concentration ratio of silver to 0.2. For the other compounds, with names HAg1:0.5, HAg1:1, HAg1:2 and HAg1:4, x values are 0.33, 0.50, 0.67 and 0.80, respectively.

## 1.2 X-ray diffraction

After sample synthesis, the X-ray diffractogram was obtained in a Shimadzu diffractometer, operating in scan mode with CuKα radiation, generated at 40 KV and 30 mA current [16]. The scanning speed used was 2/min in 2θ, with accumulation for reading every 0.6 seconds. The slits used were: divergent 1.0 mm and collection 0.3 mm. The calibration of the scanning angle was done with polycrystalline silicon and the samples were analyzed in powder form.

In the Figure 1 is shown the X-ray diffraction pattern for all $H_{1-x}Ag_xTaWO_6 \cdot nH_2O$ pyrochlore samples. We can see all peak indicate that structures belong to the Fd3m ($O_h^7$) space group, in according with reference [17], with eight molecules per unit cell. No significant structural change occurred beyond variations in relative intensities to specific peaks.

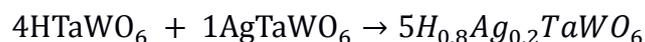

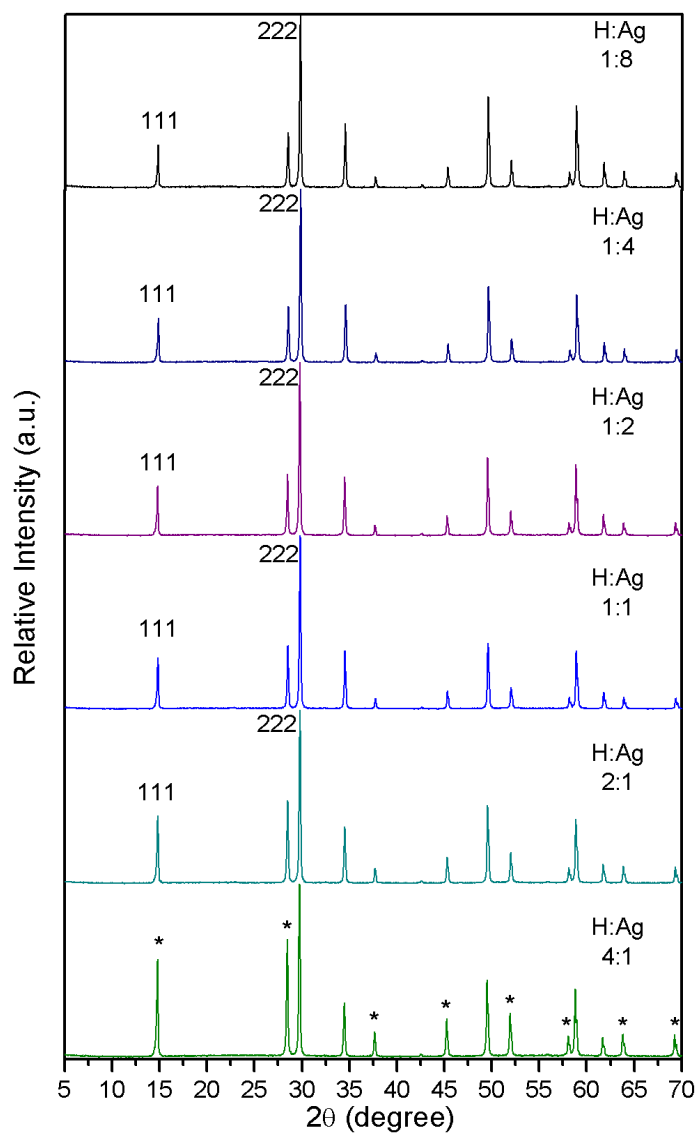

**Figure 1**: X-ray diffraction patterns for all samples synthesized in the LQES laboratory. The asterisks indicate the peaks where there are significant changes in the intensities[16].

**1.3 Impedance Spectroscopy**

The impedance spectroscopy measurements with temperature dependency were made on four samples $H_{1-x}Ag_xTaWO_6 \cdot H_2O$ pyrochlore. The experiments have performed in the laboratory of electrical measurements of the Physics Institute of the Federal University of Mato Grosso. The experimental method consisted of impedance measuring of samples within a frequency range from 1 Hz to 1MHz for each temperature, both as heating and cooling. The range of temperatures was between 25° C (27° C) and 110° C (105 or 120° C) depending upon the sample. The heating and cooling procedures have performed to study the conductivity and the activation energies for both processes.

The experimental apparatus has consisted of an impedance analyzer Solartron Analytical model 1260, Solartron Analytical 1296 dielectric interface and a commercial oven with sample holders. The diameter of the electrodes was 10 mm for samples HAg1:1 HAg1:0.5, but it was 3 mm in diameter for the other samples due to the small size of them. The thickness of each sample was 0.6 mm for the samples HAg1:0.5; HAg1:1; HAg1:4 and HAg1:0.25 and 0.65 mm for HAg1:2. As a consequence, the relation between thickness and area (electrode area) l/A was 0.076 cm$^{-1}$ for samples HAg1:1 and HAg1:0.5, 0.92 cm$^{-1}$ for HAg1:2 and 0.85 cm$^{-1}$ for samples HAg1:4 and HAg1:0.25. These values are used to determine the real and imaginary conductivities used in log-log curves of conductivity. We have obtained the conductivities (real and imaginary) from the experimental data of real and imaginary impedance through the following expressions:

$$\sigma' = \frac{1}{A} \frac{Z'}{(Z'^2 + Z''^2)} \quad (6)$$

and

$$\sigma'' = \frac{1}{A} \frac{Z''}{(Z'^2 + Z''^2)} \quad (7)$$

Where $\sigma'$ is the real conductivity, and $\sigma''$ is the imaginary conductivity. The real conductivity was the greatness used to obtain the activation energies of the pyrochlore samples.

**2 Results and Discussion**

**2.1 Activation Energies**

The parameter $\sigma_{dc}$ is thermally activated, having an Arrhenius type behavior, given by:

$$\sigma_{dc}(T) = \frac{\sigma_0}{T} \cdot exp\left(-\frac{\Delta E_a}{kT}\right), \quad (8)$$

Equation 8 describes the conductivity in the region where the defects are formed and moved, and there is a single kind of defect formed. If there is more than one type of charge carrier, this equation must be rewritten for all the different conduction mechanisms, considering its exponents and relevant pre-factors [18].

If we make manipulations in equation 8, multiplying it by T and by taking the decimal logarithm on both sides of the equation, we obtain the following equation:

$$log(\sigma T) = log(\sigma_0) - \left(\frac{\Delta E_a}{kT}\right). log(e) = log(\sigma_0) - \left(\frac{\Delta E}{1,985[eV/K]}\right)\left(\frac{10^4}{T}\right). \quad (8)$$

Where is the Arrhenius equation, valid for all models of ion conduction. The graph of log ($\sigma T$) versus $10^4/T$ is a straight line whose slope is given by $\Delta E/(1,985[eV/K])$.

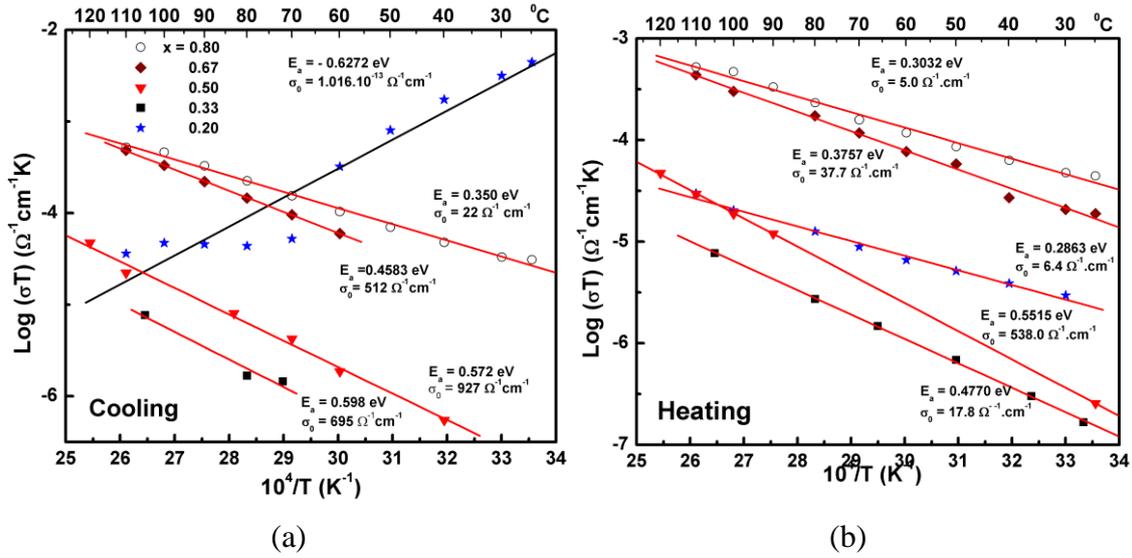

**Figure 2:** Arrhenius plot for ionic conduction for $H_{-1-x}Ag_xTaWO_6 \cdot H_2O$ samples. (a) Cooling process. (b) Heating process. Numbers represents activation energies and pre-factor $\sigma_0$ respectively.

In Figure 2 (a) and (b) it is shown the Arrhenius plots for all samples studied for the processes of heating and cooling, respectively. We have calculated the activation energies by the linear fitting of curves shown in these figures. The results of this linear fit for all samples present the following characteristics:

- The activation energy during heating for the samples were 0.30, 0.37, 0.55, 0.47 and 0.28 eV for the samples with x = (0.80, 0.67, 0.50, 0.33 and 0.20) respectively;
- The activation energies during cooling are 0.35, 0.45, 0.57 and 0.59 for the samples with x = (0.80, 0.67, 0.50 and 0.33), respectively;
- The negative curve slope (-0.62) for x = 0.20 shows that a thermally activated process did not occur. It is an abnormality that occurs in a hygroscopic material [19,20,21,22] associated with the material hydration [19,21];
- The increased doping reduces the activation energy during the heating process;
- During cooling, the samples had greater activation energies than those during heating;
- The differences in activation energy between the processes of heating and cooling for each sample are 0.05, 0.08, 0.02 and 0.12 for samples with x = (0.80, 0.67, 0.50 and 0.33), respectively. The biggest difference in activation energy is in the sample with x = 0.33;
- During both heating and cooling the activation energies for these samples are within the range of values reported in Reference [23] for the pure compound $HTaWO_6$ between hydrated and anhydrous forms (0.27 and 0.66 eV). It is a strong indication of the dependence of activation energy with doping.

By carefully examining the Arrhenius curve of the sample with rate concentration x = 0.20, we note that there is a discontinuity in it between the temperatures of 70-60 °C, where the points form two lines above and below the temperature of 70 °C. The region from 70 to 100 °C is where the water loss process by grains boundary surface dominates as seen in Chapter 1 of Reference [24]. Below this region is the reverse process: water condensation. We have, therefore, a strong evidence water condensation effect in the grain boundary.

## 2.2 Characterization of microstructure

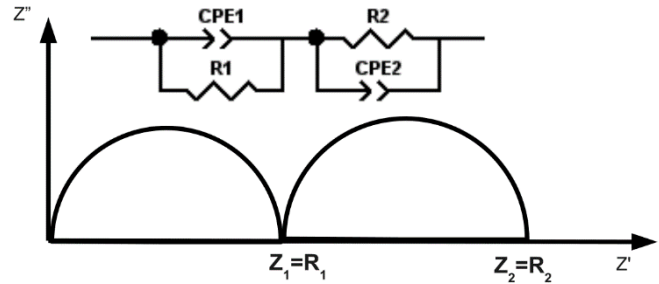

**Figure 3:** The model used for characterizing x = 0.20 sample, showing both complex impedance plot and equivalent circuit.

Figure 4 (1) and (2) shows the effect of grains and grain boundary for the samples with x = 0.80 (heating and cooling) and 0.20 during cooling, respectively. Since the time constants for each conduction process (grain and grain boundary) are very close, in sample x = 0.80 it only appears one impedance arc, as for the module, it seems a superposition of two arcs. For the sample with x = 0.20, the situation is the opposite. This situation is analogous to that shown in reference [10], where the author highlights the investigation in both impedance arcs and the electrical module to characterize materials by its microstructure. This effect also appears for the sample with x = 0.67 but not for the samples with x = 0.50 to 0.33 where there is no resolution in grain boundary (see Figure 5).

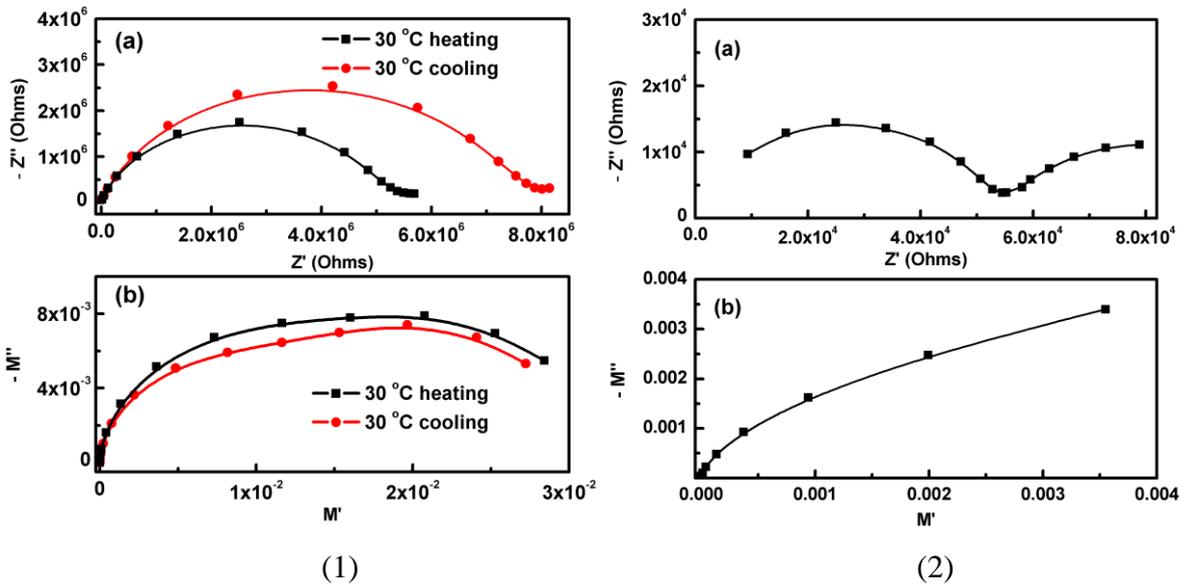

(1)                    (2)

**Figure 4:** Impedance (a) and electric Modulus (b) for (1) x = 0.80 and (2) 0.20. Similar case of references [1,16].

The electrical modulus response of the sample with x = 0.80 is typical of compounds with the Suzuki type phase, which consists of the solid-state precipitates forming islands on a second stage matrix [25]. As shown in the reference, the overlapping of complex modulus arcs indicates grains larger than 1 μm. As for the sample with x = 0.20, the microstructure is the one with grains very close to each other, i.e., very narrow grain boundary.

To confirm that, we use an equivalent circuit model to characterize the impedance of the sample with x = 0.20. The model consists of two parallel lines, R-CPE connected in series, as shown in Figure 3. From this model, we made the adjustments for the complex impedance curves for the referred sample obtained during cooling between the temperatures of 110 and 25 ºC.

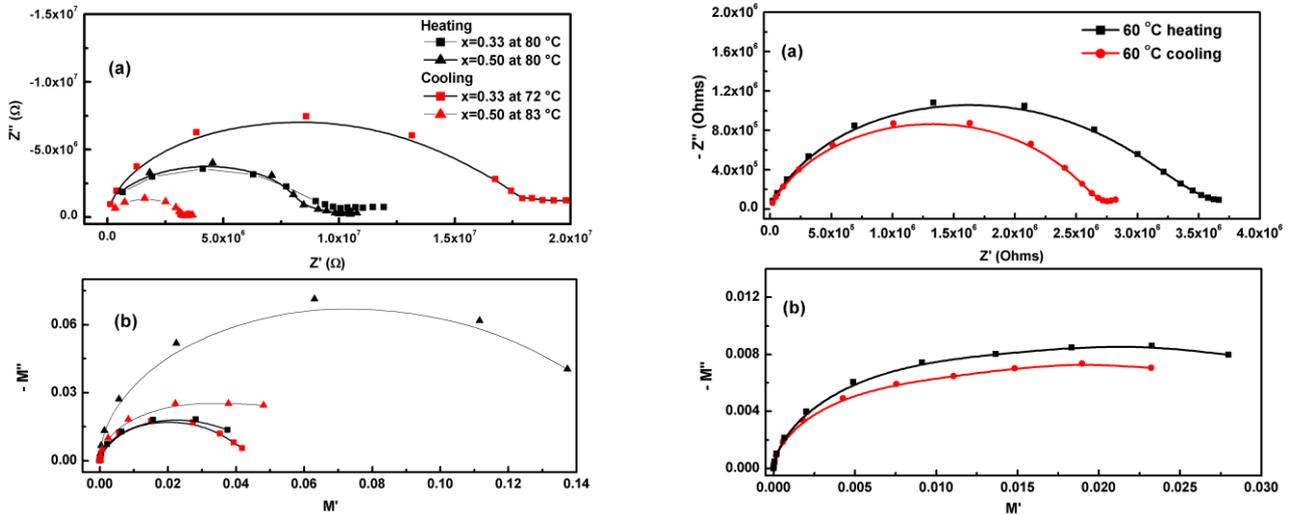

**Figure 5:** Impedance (a) and Electric Modulus (b) for (1) x = 0.50 and 0.33 and (2) x = 0.67.

The author of reference [10] in Chapter 4 of his book describes various physical models used for materials characterization. The models described range from the simplest (layers) to more complexes ones (Maxwell-Wagner). To characterize the sample with x = 0.20, we use two intermediate models: "brick-layer" and "easy-paths." The first consists of a tri-dimensional matrix of cubic grains separated by grain boundary of plane layers with a certain thickness [10], while the second consists of irregularly shaped grains where there is a region of electrical contact between them (see Figure 6).

In the "brick-layer" model, we generally assume that there is a region of continuous phase separated by individual grains. However, often where the conductivity of the grain interior is larger than the grain boundary ($\sigma_{ig} > \sigma_{cg}$) it is observed that the activation energies are equal or very similar [10]. This observation led Bauerle [26] to suggest that there are regions where the grain contours form good intergranular electrical contacts; it is called easy paths (see Figure 6).

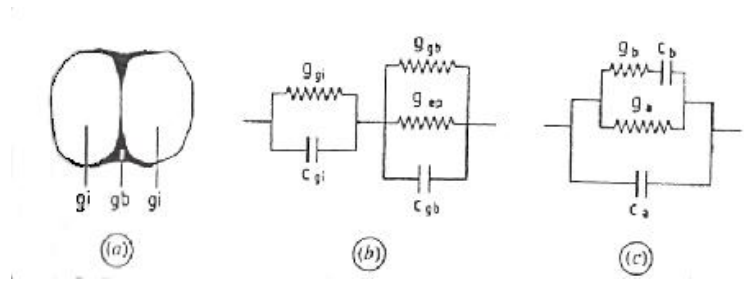

**Figure 6:** Two easy models for ceramics: (a) Schematic models representation for grain versus discontinuous grain boundaries. (b) Equivalent series circuit according Baurle(1969). (c) Parallel circuit according Schouler (1970)[10,27].

Bauerle's idea is that the migrant oxygen ions are sequentially and partially blocked in the grain's interiors and boundary. Thus there only remains the restriction of easy ionic paths (easy-paths). Since the contour permittivities and grains' interiors are equal or approximately equal ($\varepsilon_{ig} = \varepsilon_{cg}$), the existence of little intergranular contact could not noticeably affect the capacitance Ccg [10].

A slightly different model was proposed by Schouler [27]. As shown in Figure 6(c), the model divides the ionic current into two paths, one of which the conductance and capacitance (Gcg e Cig, respectively), are capacitively blocked, while another (Ga) is not. The ratio of blocked ionic current is given by

$$\beta = G_{cg}/-(G_a+G_{cg}). \tag{9}$$

Circuitry (b) and (c) of Figure 6 are equal, as shown in Chapter 1 of reference [10]. Thus, the rate β can be expressed using the equivalent circuit series

$$\beta = R_{cg}/-(R_a+R_{cg}). \tag{10}$$

When we have a situation where the activation energies for grain and grain boundary are equal (see section 3.1), the parameter β also serves to estimate the percentage with which the grain is covered by grain boundary phase, i.e., the percentage of the grain surface area that is covered by the second phase. In the case of the sample with x = 0.20 at the temperature of 100 ° C during cooling the parameter β is:

$$\beta = 0{,}83,$$

i.e., for the sample with x = 0.20, 83% of the grain surface is covered by a second grain boundary phase. This means that electrical contact between grains is only 17% of the total surface area. But this is only an estimate. The results presented here need to be confirmed through studies of the technique of Scanning Electron Microscopy.

The microstructural arrangement above, allows that a small amount of condensed water can causes a large change in the ionic conductivity of the material.

But to understand how water condensation on surfaces of grain boundary produces observable effect, it is necessary to take into account our knowledge of microstructure. The sample with x = 0.20 can have grain boundary microstructure with enough reduced volume so they can have contact points with each other (" easy path").

Figure 7 shows the Arrhenius plot of the grain resistance ($R_{ig}$) and grain boundary ($R_{cg}$) for the sample with x = 0.20, and whose values and range of temperature are shown in Table 2. The Arrhenius plots shown in the before said Figure are similar to those shown in the references [10] and [19].

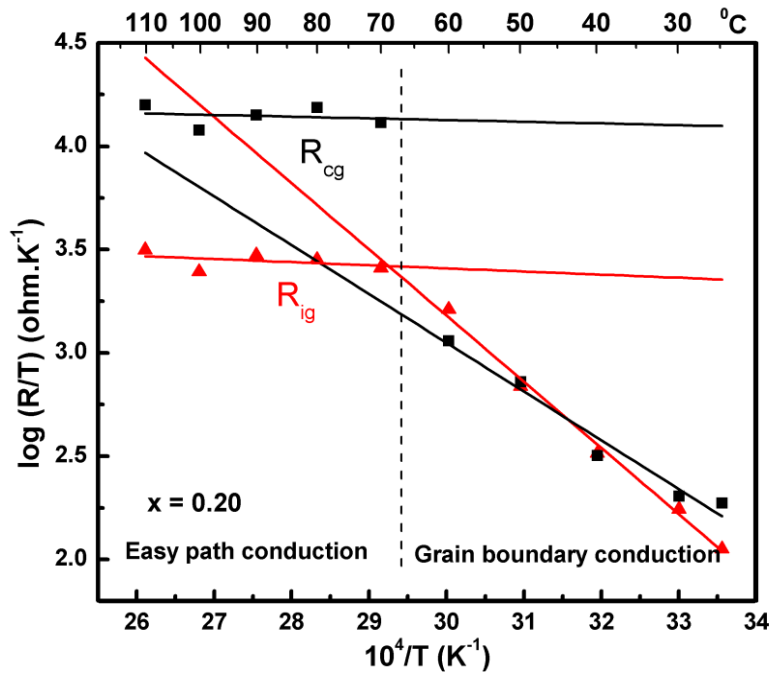

**Figure 7:** Arrhenius plot for inter and grain-boundaries resistances (x=0.20 sample). The graphic shows two regions: easy path and grain boundaries.

In the region of the conductivity of the grain boundary, the resistance grain boundary resistance Rcg is much larger than the grain interior one Rig and the angular coefficients of the straight lines are 0.32 and 0.23, therefore, very close. In the conduction region "easy path," coefficients are 0.015 and 0.008 for the resistance of interior and grain boundary, respectively. The error for each coefficient is 0.02; namely, the "activation energies" are about equal. In the previous section, we saw that Bauerle [26] used this property to explain the conduction model of "easy paths." The fact that there are points where grains have electrical contact illustrates the fact that there are activation energies of grain and grain boundary equal or similar to each other [10].

In the conduction region "easy path" the dominant thermal process is water evaporation. But as we lower the temperature, we get more water condensation, which increases the electrical contact between the grains, reducing the resistance, in an inverse process to the increasing resistance of the grains. The result is practically constant electrical resistance in this range of temperature. However, in the conduction region of the grain boundary, there is a process wherein the grain boundary dominates ionic conduction of the material. In this region, as well as the other one, it does not make sense to talk about activation energies, once it coincides with the ones that dominate water condensation. It means that the more we lower the temperature, the more surface water will be in the material, increasing conductance. Therefore, the "activation energy" for the sample with x = 0.20 appears negative, when in reality it is not occurring a thermally activated ionic conduction process in it. What occurs is that water condensation on the grains' surface eventually produces conductive paths percolation in a much faster process than the one with a decrease in conductivity of the grains.

In Figure 8, we can see the activation energy reduction with the increase of doping during the cooling process.

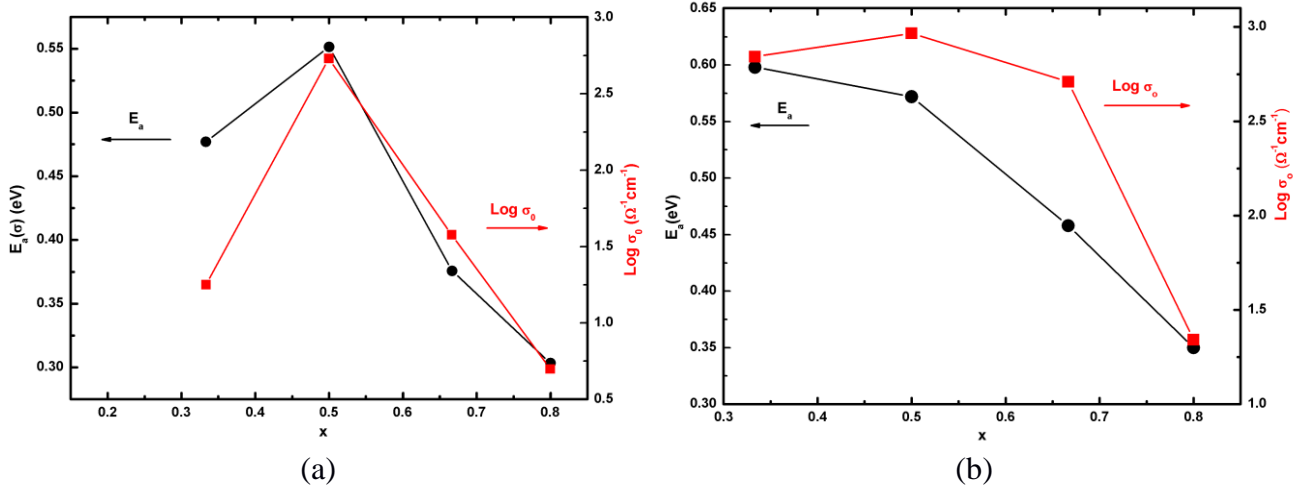

**Figure 8:** Activation energy and Log$\sigma_0$ versus x plot for both (a) heating and (b) cooling route.

These results indicate that the material HTaWO$_6$.nH$_2$O doped with silver, whose formula becomes H$_{1-x}$Ag$_x$TaWO$_6$.nH$_2$O, has as an ion conductor the H$^+$ and not the Ag$^+$. The results discussed in Chapter 3 of Reference [24] indicate that the Ag$^+$ cannot be the carrier; otherwise, the material would degrade in the TaWO$_{5.5}$ compound. Besides, among the various conduction mechanisms cited in Reference [24]: jumps, tunneling, mixed (jumps and tunneling) [23], Grothus effect [29] and the effect of surface water in the grain boundaries [29], only jump conduction and conductivity by surface water of grain boundary are possible in our samples.

The introduction of ion Ag$^+$ on 16c site weakens the bonds of octahedral Ta-O and W-O that surround the structure, reducing the vibration frequencies of the octahedral [24], which as a result, cause an enlargement of the potential wells existents within the channels where the hydrogen atoms are localized. This describing take into account the hard-core atomic model. Although it seems very unrealistic, we note in Reference [23] that an ion exchange from site 16c to another one with larger ionic ratio increases the unit cell volume and thus facilitate the conductivity. The authors showed that for a homovalent cation replacement Gd$^{3+}$ for Er$^{3+}$, Y$^{3+}$, Dy$^{3+}$, Sm$^{3+}$, and La$^{3+}$, ion conductivity decreases monotonically as the average size of the cation increases[23]. The authors also demonstrate that the material having less distension ("stress") produced by the doping atom is the one that has the highest conductivity. In our case, we found that the sample with x = 0.80, one of the samples with the lowest disorder, is the sample that has higher activation energy.

The decrease in activation energy with the doping increase is related to a compromise between two opposing processes. The first is the enlargement of the channels formed by octahedral in the directions [110] caused by the presence of Ag$^+$ ions, and the second is the blockage exercised by them. The silver ions do not take part in the conduction, as it has been seen, blocking proton conduction. Thus, the probability of an H$^+$ ion to find an Ag$^+$ ion blocking its path, restricting the possible ways that can be traversed in the structure increases with the concentration rate of Ag$^+$.

In a pure material (HTaWO$_6$), the most probable paths for conducting protons are all those that pass through the parallel channels the directions [110], which for a face-centered cubic structure are all those six directions that pass through the diagonal edges of a cube. In the case of the sample with silver rate concentration equal to 0.80, the probability of an ion H$^+$ in the center of the cube to find as a first neighbor a second ion H$^+$ or the vacancy of it (which would allow conduction) is 0.2x0.2, which is 0.04, i.e., 4%. This last one, on the other hand, has an equal probability of finding another $H^+$ ion, and this ion has 0.04 probability of finding another one, and so on, making a random path between the $Ag^+$ ion similar to a labyrinth.

The problem that relates possible random ways on two or three-dimensional networks is the percolation, which has established a minimum lower limit on the probability of a physical being finding a full path in a finite network, which depends on the freedom degrees and the used geometry. In the case of a face-centered cubic structure, such as the pyrochlore ones, the minimum percolation limit is 0.2 [30], meaning that the probability of an $H^+$ ion to find another in 0.2 is the minimum limit so percolation can happen. So, for silver concentration rate of 0.80, we are almost at the limit of optimization of the ionic conductivity. It means that for the $Ag^+$ ion as doping, we can increase the concentration rate to 0.90.

## 3 CONCLUSION

The activation energies during cooling are within lower and upper bounds for pure samples, in their anhydrous (0.60 eV) and hydrated (0.23 eV) forms. This result indicates that doping with silver influence in decreasing the activation energy.

The doping of the compound $HTaWO_6$ with $Ag^+$ increases the ionic conductivity. The explanation is that the presence of $Ag^+$ Ion extends the channels around the $H^+$ ions, facilitating the conduction [23]. The sample with x = 0.20 showed anomalies during cooling. The results plotted in the low-temperature region are similar to that reported in the literature by several authors [19-22]. The microstructure characterization indicates a possible structure "easy-paths" type, which needs to be confirmed by further studies of Scanning Electron Microscopy. For this sample, it is also required a study in Impedance Spectroscopy with a controlled atmosphere temperature.

## 4 ACKNOWLEDGMENTS

D.V. and A.G.S.F. acknowledge financial support from the Brazilian agencies CNPq and FUNCAP. We would like to thank Dr. R. J. Ramos (UFMT) for laboratory facilities, for Impedance Spectroscopy measurements.

## REFERENCES

[1] S. C. Chelgani et. al, Minerals Engineering **55**, 165–171 (2014).

[2] D. GROULT, B. RAVEAU, and C. MICHEL, COMPTES RENDUS HEBDOMADAIRES DES SEANCES DE L ACADEMIE DES SCIENCES SERIE C **274**, 374.

[3] M.A. Subramanian, G. Aravamudan, and G.V.S. Rao, Prog. Solid State Chem. **15** 55–143 (1983).

[4] M.A. Subramanian, B.H. Toby, A.P. Ramirez, W.J. Marshall, A.W. Sleight, and G.H. Kwei, Science 273 (1996) 81–84.

[5] B.J. Kennedy, T. Vogt, J. Solid State Chem. 126 (1996) 261–270.

[6] S. Yonezawa, Y. Muraoka, Y. Matsushita, Z. Hiroi, J. Phys.: Condens. Matter 16 (2004) L9–L12.

[7] C. M. Mari, F. Bonino, M. Catti, R. Pasinetti, and S. Pizzini, Sol. Stat. Ion. **18-19**, 1013 (1986).

[8] J. Hidalgo, M. Colet-Lagrille, and A. Mukasyan, Cryst. Res. Technol. **50**, No. 11, 879–883 (2015).

[9] M. Catti, C. Mari, and G. Valerio, Journ. Sol. Stat. Chem. **98**, 269 (1992).

[10] J. R. Macdonald, *Impedance Spectroscopy Emphasizing Solid Materials and Sistems*, JOHN WILEY SONS, 1 edition, 1987.

[11] C. Leon, M. L. Lucia, and J. Santamaria, Physical Review B **55**, 882 (1997).

[12] C. M. Mari, M. Catti, and M. Castelli, A Characterization of $KTaWO_6 \cdot H_2O$. **Mat. Res. Bull.,** v.21. p. 773-778, (1986).

[13] Zarbin, Aldo J G, Alves, Oswaldo L, Amarilla, J Manuel, Rojas, R M, and Rojo, J.M, **Chem. Matter.** V. 11, p. 1652-1658, (1999).

[14] A. Galembeck, and O.L. Alves. J. Mater. Sci, v.34, p. 3275-3280, (1999).

[15] J A Díaz-Guillén, A F Fuentes, M R Díaz-Guillén, J M Almanza, J Santamaría, and C Léon, Journal of Power Sources **186**, 349 (2009).

[16] M. A. C. de Santis, Estrutura lamelar versus estrutura pirocloro: Obtenção de compostos do tipo $H_{1-x}Ag_xTaWO_6$, Master thesis, UNICAMP-Universidade Estadual de Campinas, 2006.

[17] C. M. Mari, M. Catti, and M. Castelli, A Characterization of $KTaWO_6 \cdot H_2O$. **Mat. Res. Bull.,** v.21. p. 773-778, (1986).

[18] E. N. Silva, Espectroscopia vibracional e de impedância em monocristais de $SrAlF_5$, Master thesis, Universidade Federal do Ceará, 2005.

[19] C. M. Mari, F. Bonino, M. Catti, R. Pasinetti, and S. Pizzini, Sol. Stat. Ion. **18-19**, 1013 (1986).

[20] C. M. Mari, M. Catti, A. Castelli, and F. Bonino, Mat. Res. Bull. **21**, 773 (1986).

[21] M. Catti, E. Cazzanelli, C. M. Mari, and G. Mariotto, J. Solid State Chem. **107**, 108 (1993).

[22] M. Catti and C. M. Mari, Sol. Stat. Ion. **40/41**, 900 (1990).

[23] M.A. Butler and R. M. Biefeld, Phys. Rev. B **19**, 5455 (1979).

[24] D. Valim, *Espectroscopia vibracional e de impedância de $A_{1-x}A`_xTaWO_6 \cdot nH_2O$ (A=H, Li, A'=Ag, H)*, PhD thesis, Universidade Federal do Ceará, Fortaleza-CE, 2009.

[25] N. Bonanos and E. Lilley, J. Chem. Solids **42**, 943 (1981).

[26] J. E. Bauerle, Journal of Physical Chemistry Solids **30**, 2657 (1969).

[27] E. J. L. Schouler, *Etude des Cellules a Oxyde Electrolyte Solide per la Methode des Impedances Complexes*, Ph.d thesis, Institut National Potytechinique de Grenoble, 2005.

[28] E. Lilley and J. E. Strutt, Physics State Solid **54**, 639 (1979).

[29] M. Catti, C. Mari, and G. Valerio, Journ. Sol. Stat. Chem. **98**, 269 (1992).

[30] C. D. Lorentz and R. M. Ziff, Journal of Physics A **31**, 8147 (1998).